\documentclass[10pt,letterpaper,twocolumn]{article}
\usepackage{fullpage}

\newcommand{\pc}{\text{pc}}
\newcommand{\kpc}{\text{kpc}}

\usepackage[normalem]{ulem}
\usepackage{xcolor}
\makeatletter
\def\squiggly{\bgroup \markoverwith{\textcolor{red}{\lower3.5\p@\hbox{\sixly \char58}}}\ULon}
\makeatother

\usepackage{subfig}
\usepackage{amsmath}
\usepackage{amssymb}
\usepackage{graphicx}
\graphicspath{{./fig/}}

\begin{document}

\title{Dissipative Dark Matter and the Andromeda Plane of Satellites}
\author{Lisa Randall and Jakub Scholtz\\{\small Department of Physics, Harvard University, Cambridge, MA 02138, USA}}

\twocolumn[
  \begin{@twocolumnfalse}
    \maketitle
    \begin{abstract}
      We show that  dissipative dark matter can potentially explain the large observed mass to light ratio of the dwarf satellite galaxies that have been  observed in the recently identified planar structure  around Andromeda, which are thought to result from tidal forces during a galaxy merger.  Whereas dwarf galaxies created from ordinary disks would be dark matter poor, dark matter inside the galactic plane not only provides a source of dark matter, but one that is more readily bound due to the dark matter's lower velocity.  This initial  N-body study shows that with a thin disk of dark matter inside the baryonic disk, mass-to-light ratios as high as $\mathcal{O}(30)$ can be generated when tidal forces pull out patches of sizes similar to the scales of Toomre instabilities of the dark disk. A full simulation will be needed to confirm this result.\\  
    \end{abstract}
  \end{@twocolumnfalse}
]

\section{Introduction}

Recent observations of dwarf satellites in the Andromeda galaxy \cite{Ibata:2013rh} show them to have a distribution with about half (15 out of 27) in a single plane about 14 kpc thick and 400 kpc in radius. Furthermore, most of these (13 out of 15) are rotating in the same direction. This surprising result contradicts the expectation from usual cold dark matter (CDM) models,  which tend to  predict a more isotropic distribution. Similar claims have been made about the Milky Way \cite{2014ApJ79074P}, which also exhibits more planar dwarf galaxy distribution than expected.  Furthermore, other even more recent results show such a  planar distribution of satellite galaxies to be surprisingly common, with many galaxies located in a plane \cite{2014Natur511563I}, although \cite{2014arXiv1410.7778C} cautions that this claim may be too sensitive to the selection criteria for the galaxies investigated. 
 
\begin{figure}%
\centering
\includegraphics[width=\columnwidth]{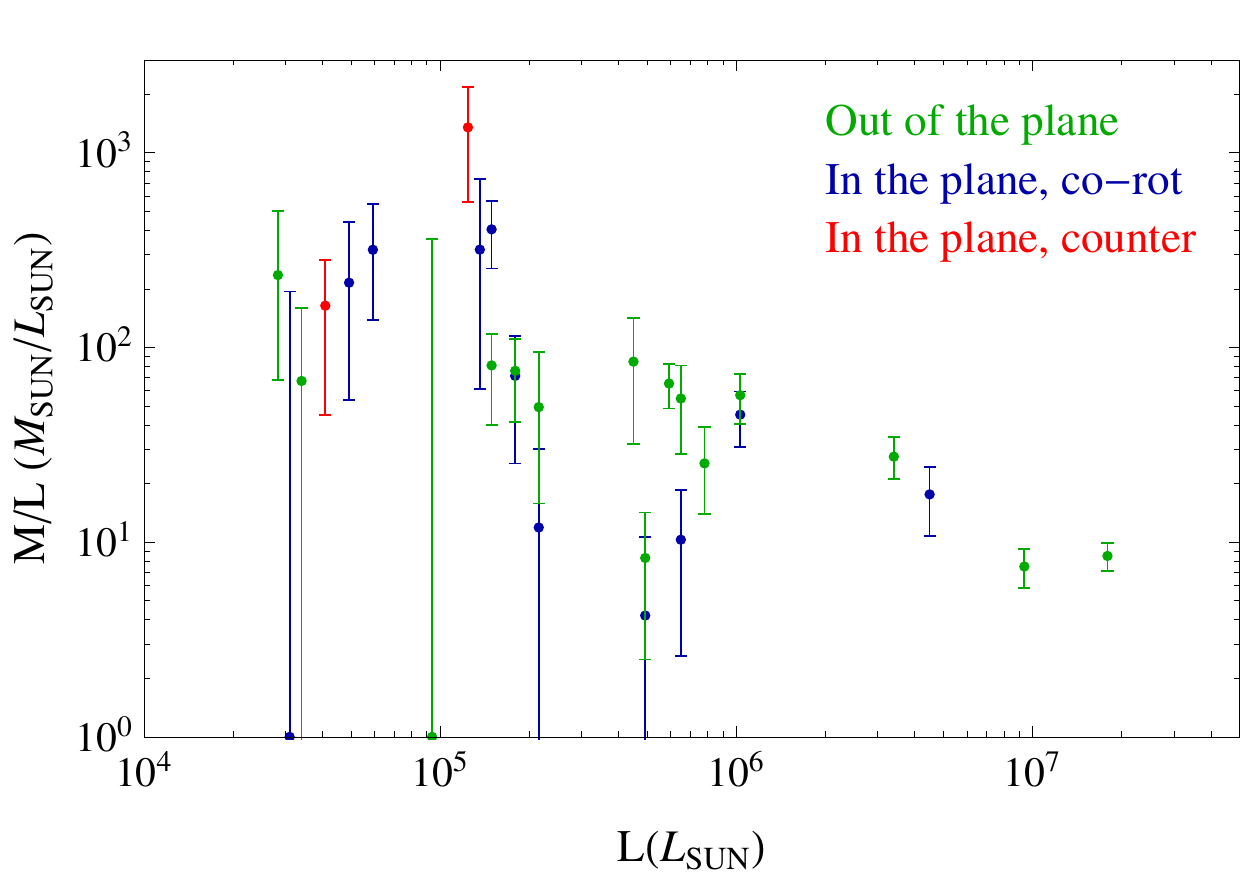}%
\caption{Comparison between out-of-the-plane (green), co-rotating in-the-plane (blue) and counter-rotating in-the-plane (red) dwarf satellites of the Andromeda Galaxy. Neither population appears distinct in terms of their $M/L$ values. The $M/L$ values and error bars have been obtained by applying the method from \cite{2012AJ1444M} to data from Table~5 of \cite{Collins:2013swa} and therefore agree with column $M/L(3)$ of table~\ref{tab:mol} in this work. }%
\label{fig:MoLvsL}%
\end{figure}

Of course, this surprising result can potentially tell us about non-standard phenomena only if standard explanations do not successfully address the data. Galactic dynamics can certainly modify spatial distributions from naive expectations.  Indeed, tidal forces arising from merging galaxies might well account for the spatial distribution  \cite{Hammer:2010ug,Hammer:2013bga}.  However, even if the dynamics of tidal stripping explains the spatial structure, a puzzle still remains. Studies  \cite{2012AJ1444M,Collins:2013swa} show evidence for large mass to light ($M/L$) ratios in these satellites; we have summarized the results of these studies in table~\ref{tab:mol} and figure~\ref{fig:MoLvsL}. Although these $M/L$ values come with large uncertainties, figure~\ref{fig:MoLvsL} shows there is a consistent trend indicating many of these dwarf galaxies are dominated by dark matter. Also, at this point, there is no obvious difference between the in-the-plane and the out-of-the-plane satellites in terms of their $M/L$ values. Both in-the-plane and out-of-the-plane populations display a range of $M/L$ values -- and our model will result in a similar range of values for the in-the-plane satellites.
\begin{table}
\centering
\begin{tabular}{l|lll}
\hline 
dSph & $M/L(1)$ & $M/L(2)$ & $M/L(3)$\\
\hline
And I & $21\pm 5$ & -- & $18\pm 7$ \\ 
And III & $14\pm 12$ & -- & $45\pm 14$ \\ 
And IX & $88^{+167}_{-88}$ & -- & $404 \pm 150$ \\ 
And XI & $78^{+183}_{-78}$ & $216^{+115}_{-87}$ & $215\pm 162$ \\ 
And XII & $77^{+176}_{-77}$ & $0^{+194}$ & $0^{+194}$ \\ 
And XIV & $63 \pm 55$ & -- & $71 \pm 46$ \\ 
And XVI & $39^{+79}_{-39}$ & -- & $4.2^{+6.4}_{-4.2}$ \\ 
And XVII & -- & -- &  $12^{+16}_{-12}$\\ 
And XXV & -- & $10.3^{+7.0}_{-6.7}$ & $10 \pm 8$ \\ 
And XXVI & -- & $325^{+243}_{-225}$ & $318 \pm 179$ \\ 
Cass II & -- & $308^{+269}_{-219}$ & $318 \pm 257 $ \\ 
NGC 147 & $3\pm 0.5$ & -- & -- \\ 
NGC 185 & $4\pm 0.5$ & -- & -- \\ 
\hline 
\end{tabular} 
\caption{Collected values of $M/L$ for co-rotating in-the-plane dwarf satellites of the Andromeda Galaxy. The $M/L(1)$ values come from \cite{2012AJ1444M}, we have computed the $M/L$ error bars from their data. The $M/L(2)$ values and error bars are from Table~4 of \cite{Collins:2013swa}. We have also used the same method as \cite{2012AJ1444M} to compute the $M/L(3)$ values and error bars based on data from Table~5 of \cite{Collins:2013swa}.}
\label{tab:mol}
\end{table}

Such large $M/L$ values cannot be accounted for by just baryonic matter and are therefore inconsistent if tidal forces are responsible for creation of these dwarf satellites. According to conventional scenarios, not enough ordinary cold dark matter would be trapped in these structures to agree with measured mass to light ratios since tidal stripping does not incorporate CDM. Unless an explanation for the sizable amount of nonluminous matter contained in these dwarfs, the results will remain perplexing.

In fact, there is no known universally accepted resolution to this puzzle. Yang et al \cite{Yang:2014yia} propose for the Milky Way Galaxy that, over time, satellites absorb dark matter and lose baryons, thereby raising the mass to light ratio. However, this resolution is specific to the Milky Way satellites and it is not clear how applicable it is to other systems.  Another proposed explanation is that there is simply a good deal of nonluminous ordinary matter. But no evidence to support this hypothesis is observed. An alternative explanation was put forth by Foot et al. \cite{Foot:2013nea}, who suggest that a disk of mirror dark matter might play a role, though they have no explicit mechanism or dynamics for how this came about. 

Given our limited understanding of dark matter, results such as these are worthy of attention. If dark matter continues to elude detection, it could be that exploring matter spatial distributions throughout the universe will give us best clues as to whether dark matter has interactions other than gravitational.

In this paper, we propose an admittedly speculative idea based on the Partially Interacting Dark Matter (PIDM) scenario proposed by Fan et al. \cite{Fan:2013yva}. In PIDM scenarios most dark matter is non-interacting or extremely weakly interacting and resides in a smooth spherical halo, but a small fraction of the dark matter has self-interactions. Double Disk Dark Matter (DDDM) includes dissipative self-interactions in the dark sector. If the dark matter particle is heavier than the proton the DDDM forms a thin disk of dark matter embedded in the plane of the galactic disk. We propose that tidal forces acting on such a disk, with a thicker disk of baryonic matter and a thinner disk containing a small fraction of dark matter, could account for observations of planes of Tidal Dwarf Galaxies with large mass-luminosity ratio. Although the dissipative dark matter constitutes only a fraction of the mass of the initial disk, a thin disk is indicative of a dark matter mass heavier than that of the proton as well as a velocity lower than that for baryons. Because of the lower initial velocity and the lack of interaction with the more energetic baryons, dark matter particles are more likely to be trapped into bound structures when tidal forces pull them out, and hence large mass to light ratios are expected.

\subsection{PIDM}
\label{sec:PIDM}
Reference~\cite{Fan:2013yva} proposes that the dark matter is composed of at least two different sectors. The majority of dark matter (DM)  has all the assumed properties of CDM -- it does not interact very much with either the Standard Model or with itself. However, a small portion of the dark matter is self-interacting and dissipative. A particularly simple model consists of two fermions $X$ and $C$ charged under a $U(1)'$ that only mediates forces among dark sector particles. The assumed mass of $X$ is bigger than proton mass, $m_X > 1\;\text{GeV}$, while $C$ is light: $m_C \lesssim m_e$. The $U(1)'$ vector (the dark photon) is massless and therefore offers the possibility of carrying away arbitrarily small amounts of energy and so both $X$ and $C$ are able to dissipate energy into the dark radiation. Whereas the fermion $X$ constitutes majority of the energy density of the PIDM, fermion $C$ acts a coolant because its Thompson scattering cross-section with dark photons is much larger:
\begin{equation}
\sigma_T = \frac{8\pi}{3}\frac{\alpha_D^2}{m_C^2}
\end{equation}
We will work with a scenario in which a galaxy contains equal number of oppositely charged $X$ and $C$ particles so that at sufficiently low temperatures $X$ and $C$ should combine into a hydrogen-like bound state with binding energy:
\begin{equation}
B = \frac{\alpha_D^2 m_C}{2}
\end{equation}
When the temperature of the dark sector drops below $T \sim B/20$ we expect the majority of the gas  to recombine \cite{Fan:2013yva}. Indeed, we assume that the gas of unbound $X$s and $C$s continues cooling via bremsstrahlung (and initially via Compton scattering on dark radiation) until it reaches this temperature, below which the cooling essentially stops.

The recombination temperature, $T$, can be comparable or even lower than the temperatures available to the baryonic sector. If the $X$ particle is heavier than the proton, this would imply a thinner disk for the interacting component of dark matter. In fact, for a common binding energy, the thickness of the disk scales inversely with the mass of $X$ and so a PIDM disk that is more narrow than the visible galactic plane should form. This disk is very likely aligned with the baryonic disk due to their mutual gravitational attraction as well as common tidal torques acting on both disks. However, because of the $X$ particle mass, we expect the dark matter disk to be narrower than the disk of ordinary matter.  Moreover, whatever pressure support the baryonic disk enjoys, the PIDM disk may lack.  This means that dynamics for the dark disk can be complicated. Depending on the local surface densities and $T$, the PIDM disk can be unstable and form clumps of characteristic size as described in section~\ref{sec:clumps}.

\section{Methodology}

\subsection{Basic Dynamics}
\label{sec:basicdyn}
Our analysis uses as a starting point the work of Hammer et al., who propose that Andromeda is the result of at least one merger of two large objects whose masses differ by a factor of about three. If this is the case, tidal forces will likely affect the overall structure of the galaxy, ejecting some material from its progenitors that can remain bound into the overall structure. Using numerical simulations, the authors of \cite{Hammer:2013bga} have shown they can closely model the size and spatial distribution of the observed tidal streams, which they then assume can fragment into the observed dwarf galaxies. With appropriate simulation parameters, their explanation seems to match the satellites' planar distribution well.  However, they cannot explain the large amount of nonluminous matter these satellites contain.

We apply the Hammer mechanism to our scenario with progenitor galaxies that contain a thicker disk of baryonic matter and a thinner disk of PIDM embedded inside. As in the Hammer scenario, tidal forces pull out large amounts of matter from the disk, in our case both baryonic and dark. This is typically not the case in the ordinary CDM scenario, because the total amount of CDM in the galactic disk is insignificant. In our scenario too, initially, the amount of baryonic matter far exceeds the amount of dark matter that is pulled out of the disk, with baryons dominating by roughly a factor of five or six.

However, a significant feature of PIDM is that the dispersion velocity of the dark matter can be much lower than that of baryons. Therefore, when tidal forces pull both baryons and dark matter from the galactic plane, dark matter particles are much more likely to remain bound into an emergent structure while baryons are more likely to evaporate. Due to the fact that the process of evaporation of unbound baryons and PIDM is out of equilibrium, highly time dependent, and  strongly influenced by energy transfer between the particles, we currently cannot analytically predict the amount of baryons and PIDM that stay bound. We instead perform numerical simulations to evaluate the fraction of baryons that might remain. 

Our goal in this paper is to show that given tidal disruption, a large fraction of dark matter will  remain bound in any dwarf satellite galaxies that might form. Therefore, in our simulations, we assume tidal forces pull out a patch of material from the original galaxy, taking the tidal disruption as a given and focusing solely on the patch and how it evolves with no other matter around it. The size of the patch can be determined by the scales on which the tidal forces operate but also by other scales such as the size of gravitationally unstable objects in the dark disk. The patch cannot be much smaller than these unstable regions, because once these instabilities form they are much harder to split. On the other hand, if this patch becomes much bigger than the size of gravitational instabilities we can separate the behavior into evolution of each gravitational instability and further interaction of these instabilities. We will describe these gravitational instabilities, clumps, in section~\ref{sec:clumps}.  Ideally a full simulation would model the gravitational instabilities and the tidal disruption and determine the characteristic scales for us. Our intention in this paper is only to show that PIDM leads to a much larger fraction of dark matter than the usual CDM scenario.

A typical initial state has significantly more baryons. About $10-20\%$ of these would be bound initially  because their initial velocity is smaller than the escape velocity from the entire cloud. However, since baryons have much higher spatial overlap with themselves than with the PIDM, a lot of these baryons gain enough energy to become free. The PIDM particles on the other hand, start off with lower velocity and are furthermore less likely to interact with the higher velocity  baryons and hence do not evaporate as much. Based on these simulations, we find this to be the case, with the final dark matter mass often exceeding that in the baryons by at least a factor of 10.

\subsection{Progenitor Galaxies}
\label{sec:progenitor}
Since the smaller progenitor galaxy (SPG) is more likely to get tidally disrupted, we assume  that the patch that eventually forms the tidal dwarfs comes from the SPG. In order to characterize this patch we need to know the SPG's surface density, its local density at the disrupted region and hence its scale radius, the local orbital velocity that tells us both the stabilizing angular momentum and which also contributes to the total kinetic energy in the patch, as well as the dispersion velocities for both the PIDM and for the baryons.  The dispersion velocity of the PIDM follows from the model, as shown in section~\ref{sec:PIDM}. The remaining parameters can be estimated by rescaling the values from the Milky Way galaxy to the mass of the smaller of the progenitors estimated by \cite{Hammer:2010ug}. Hammer et al. argue that Andromeda is a result of a merger of two galaxies with mass ratio of about three, which makes the SPG's mass about $M_{\text{Andromeda}}/4$. We use the Tully-Fisher relationship $L \propto v^4$ and assume that the mass to light ratio is similar to rescale the Milky Way properties in order to determine the SPG properties similarly to what was done in~\cite{Barnes:2002sh}.

By rescaling the mass and velocity according to the Tully-Fisher relationship, we find the best estimate for radial disk scale to be half that of the Milky Way, or $r_d = 1.5\kpc$. Using the surface density at the Sun's location to normalize and applying the above scale length yields at the center of the SPG $\Sigma_0 = 860 M_\odot/\pc^2$ (which corresponds to $60 M_\odot/\pc^2$ in the solar neighborhood). The rotation curve flattens out at $R = 2\kpc$ and reaches the asymptotic velocity $v_0 \sim 165\text{km/s}$.
 
Since the mass of Andromeda is uncertain,  we also redo our analysis assuming Andromeda is larger by a factor of 1.5. In that case the mass of the SPG  is that of Milky Way rescaled by a factor of $1.5/4 = 0.375$. This means the SPG radial scale is $r_d = 1.84\kpc$ and the orbital velocity in the flat rotation curve goes to $v_0 \sim 183\text{km/s}$. 

\subsection{Clumps}
\label{sec:clumps}

\begin{figure}%
\centering
\includegraphics[width=0.8\columnwidth]{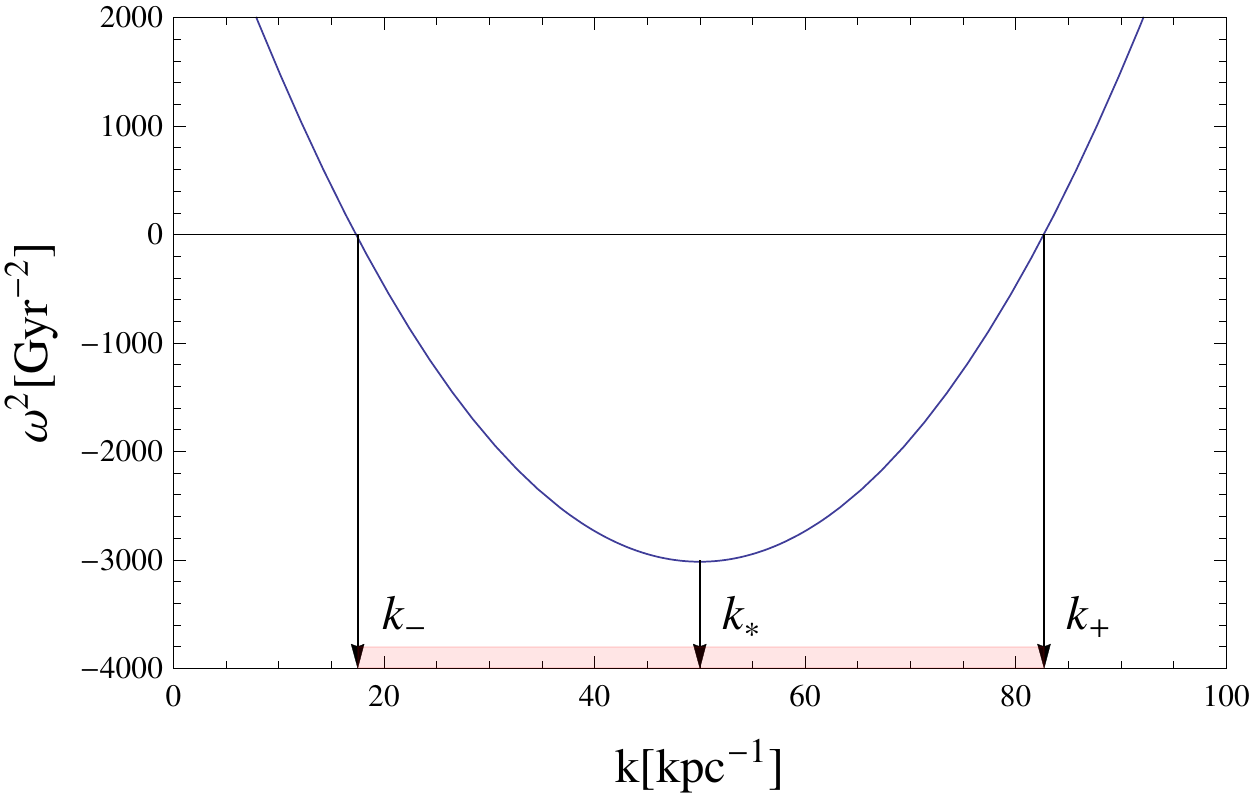}%
\caption{Illustration of the Toomre Criterion with parameters that correspond to local properties of SPG at $R=4\kpc$. The red marked range of $k$s corresponds to density perturbations that grow with time. The largest and smallest $k$ modes that exhibit growing behavior are marked $k_-$ and $k_+$, while the mode that initially grows the fastest is marked $k_*$.}%
\label{fig:till}%
\end{figure}

As we argued in section~\ref{sec:basicdyn}, each patch consists of one or more gravitational instabilities, clumps. Both baryons and PIDM are subject to gravitational instabilities that may cause clumping in the original smooth distribution of matter within the disks. In a uniform three dimensional distribution the criterion for existence of gravitational instabilities is the Jeans mass. The Toomre criterion is the analog of the Jeans mass in a two dimensional system with angular momentum,  such as disks.  The Toomre criterion is a good indicator of the onset of  instabilities as it describes the time evolution of an amplitude of a small density perturbation with a particular wavelength. When $\omega(k)$, the oscillation frequency of the amplitude of a density perturbation with wavelength $k$, gains a non-zero imaginary component, these density perturbations grow (and quickly violate the linear assumption used to derive this expression). The expression for $\omega(k)$ is simple:
\begin{equation}
\kappa^2 - 2\pi G \Sigma(R) |k| + \sigma^2 k^2  = \omega(k)^2 < 0,
\end{equation}
where $\kappa = \sqrt{2}\Omega$ is the epicyclic frequency of the disk ($\sqrt{2}$ corresponds to a flat rotation curve), $\sigma$ is the velocity dispersion of the matter in the disk and $\Sigma(R)$ is the local surface density of the disk. Notice that large $\kappa^2$ stabilizes the disk and so does large $\sigma^2$. In the first case,  the disk is angular momentum supported and in the latter it is pressure supported. On the other hand, the term proportional to $G\Sigma$ lowers $\omega^2$ and induces instabilities, just as we would expect. If the disk is unstable ($\omega^2(k) < 0$ for some $k$) there are three wavenumbers of interest:
\begin{align}
k_+ &= \frac{\pi G \Sigma}{\sigma^2} + \sqrt{\left(\frac{\pi G \Sigma}{\sigma^2}\right)^2- \frac{\kappa^2}{\sigma^2}}\\
k_- &= \frac{\pi G \Sigma}{\sigma^2} - \sqrt{\left(\frac{\pi G \Sigma}{\sigma^2}\right)^2- \frac{\kappa^2}{\sigma^2}}\\
k_* &= \frac{\pi G \Sigma}{\sigma^2}
\end{align}
Whereas $k_+$ and $k_-$ correspond to the smallest and largest unstable regions (for which $\omega(k_\pm) = 0$), respectively, $k_*$ is the wavenumber of the least stable perturbation ($\omega^2(k_*) = \min(\omega^2)<0$). Figure~\ref{fig:till} shows the different $k$-numbers of interest.

  Notice that $2/k_*$  is equal to the characteristic scale height of the disk $h_{D}$ (under the assumption $\sigma = \sigma_r = \sigma_z$). This means that the most unstable region is essentially the same as the region for which the three-dimensional Jeans instability applies. However, we will focus here not on the most unstable but on the largest unstable region of the dark disk. 

In our model the velocity dispersion is purely thermal and the PIDM cools down to a temperature of order $T \sim B/20 = \alpha_D^2 m_C /40$, ($B$ is the binding energy of the dark hydrogen). As a result the velocity dispersion obeys a simple relation:
\begin{equation}
\sigma = \frac{\alpha_D c}{\sqrt{40}}\sqrt{\frac{m_C}{m_X}}
\end{equation}
The velocity dispersion $\sigma$ determines the scale height of the disk $h_D$:
\begin{equation}
h_D = \frac{\sigma^2}{\pi G \Sigma_{dm}} = \frac{\alpha_D^2 c^2}{40\pi G \Sigma_{dm} }\frac{m_C}{m_X},
\end{equation}
and therefore we can, and will, treat $h_D$ as a free parameter in our study. 

In our N-body simulations we set the size of the patch pulled out of the SPG to be of the same order as the size of the largest unstable region in the dark disk. This is a self-consistent approach because all the particles from a particular clump that might end up in the same gravitationally bound object are included in the simulation. However, since we do not know the exact size of the patch that gets pulled out, we treat $x$ as a free parameter and show results for a range of $x$'s.

\subsection{Simulation}
\label{sec:simulation}
How much of the matter that gets lifted out of the disk remains bound in the tidal stream and how much of the matter flows out? The answer depends on the kinetic energy of the material that is pulled out. Because baryons under our dark disk assumption have bigger velocity than the dark matter particles, baryons evaporate out of the region at a higher rate than the dark matter. Even many of those baryons on the low velocity tail that  initially had low enough velocity to be bound will be removed by their interactions with other energetic baryons, which will kick them out of the bound system. So even though baryons dominate the initial tidally produced clump, dark matter can dominate in the end. However, it is subtle to calculate the values in this complicated dynamical system where energy is exchanged between baryons and dark matter so we choose to do a crude dynamical simulation to evaluate what will occur.

In order to determine how many particles of each type stay bound to the patch that has been pulled out by tidal forces we perform a series of N-body simulations. We replace the original smooth density distribution of matter by $N$ particles that are modeled by Plummer spheres (defined in appendix~\ref{sec:plummer}) with smoothing length $\epsilon$ of the same order as the initial interparticle spacing. We numerically integrate the equations of motion of these particles with a leapfrog algorithm, as defined in appendix~\ref{sec:leapfrog}. We chose the leapfrog algorithm for its relative simplicity and high degree of energy conservation, which is important because our analysis depends on determining the final distributions of binding energies of these particles.  The first $\lfloor N/6 \rfloor$ particles are assigned parameters to represent PIDM and the last $\lceil 5N/6 \rceil$ represent baryons. Our simulation takes the mass of each particle to be the same $m = M_{tot}/N$. At the end of the simulation we evaluate the binding energy of each particle:
\begin{equation}
E_B^i = E_k^i +E_p^i = \frac{1}{2} m v_i^2 -\sum_{i\neq j} \frac{G m^2}{|r_i-r_j|}
\end{equation}
If the binding energy $E_B$ is negative, we consider the particle bound and assume it will become part of a tidal dwarf galaxy. Conversely, if the binding energy is positive, the particle escapes and does not belong to the bound structure. We do not observe a significant number of binaries\footnote{We believe this is due to relatively large $\epsilon$.} and therefore asymptotically this binding energy represents true binding energy with respect to the remnant body. In order to test this, we also compute the potential energy with respect to the largest group of particles found by a Friends of Friends algorithm \cite{FOF}. We call these mutually disjoint groups cores. The binding energy with respect to this core is:
\begin{equation}
E_B^i(\text{core}) = \frac{1}{2} m (v_i-v_{\text{core}})^2-\sum_{j \in \text{core}} \frac{G m^2}{|r_i-r_j|},
\end{equation}
where $v_{\text{core}}$ is the velocity of center of mass of the core. We confirm these two different definitions of binding energy give us similar numbers of bound and free particles at the end of our simulation.  However, in some scenarios the final state consists of more than one self-bound object. In this case we compute the binding energies with respect to two largest cores and compare the results. In order to reduce the dependence of our numerical integration on the representation of the initial state we repeat each simulation ten times with exactly the same parameters but different random number seeds and average over the quantities extracted from these simulations.

\subsection{Initial State}

The spatial distributions of each species are uniform in the $xy-$plane inside a square of size $x$, which we take to be a range of values of order $x = 2\pi/k_-$ with densities that scale as an exponential in the z-direction with scale heights $h_{D}$ and $h_{B}$.
\begin{align}
\rho_b &= \frac{\Sigma_b (R)}{2h_B} \exp\left(-\frac{|z|}{h_B}\right)\\
\rho_{dm} &= \frac{\Sigma_{dm} (R)}{2h_{D}} \exp\left(-\frac{|z|}{h_{D}}\right)
\end{align}
The velocity distributions contain two contributions. Aside from the Boltzmann distributed velocities, there is a velocity field corresponding to the bulk rotation of the particles with a flat rotation curve with velocity $v_0$ with the center of the square displaced by $R$ from the SPG's center.
\begin{equation}
\left(\begin{matrix} v_x \\ v_y \\ v_z  \end{matrix}\right)_{Bulk} = \frac{v_0}{\sqrt{(x+R)^2+y^2}}\left(\begin{matrix} -y \\ x+R \\ 0\end{matrix}  \right)
\end{equation}
The second contribution to the velocity distributions is the Boltzmann distribution for each component with dispersion velocities $\sigma_{dm}$ and $\sigma_b$.

\subsection{Simulation Consistency Checks}

\begin{figure}
\includegraphics[width=\columnwidth]{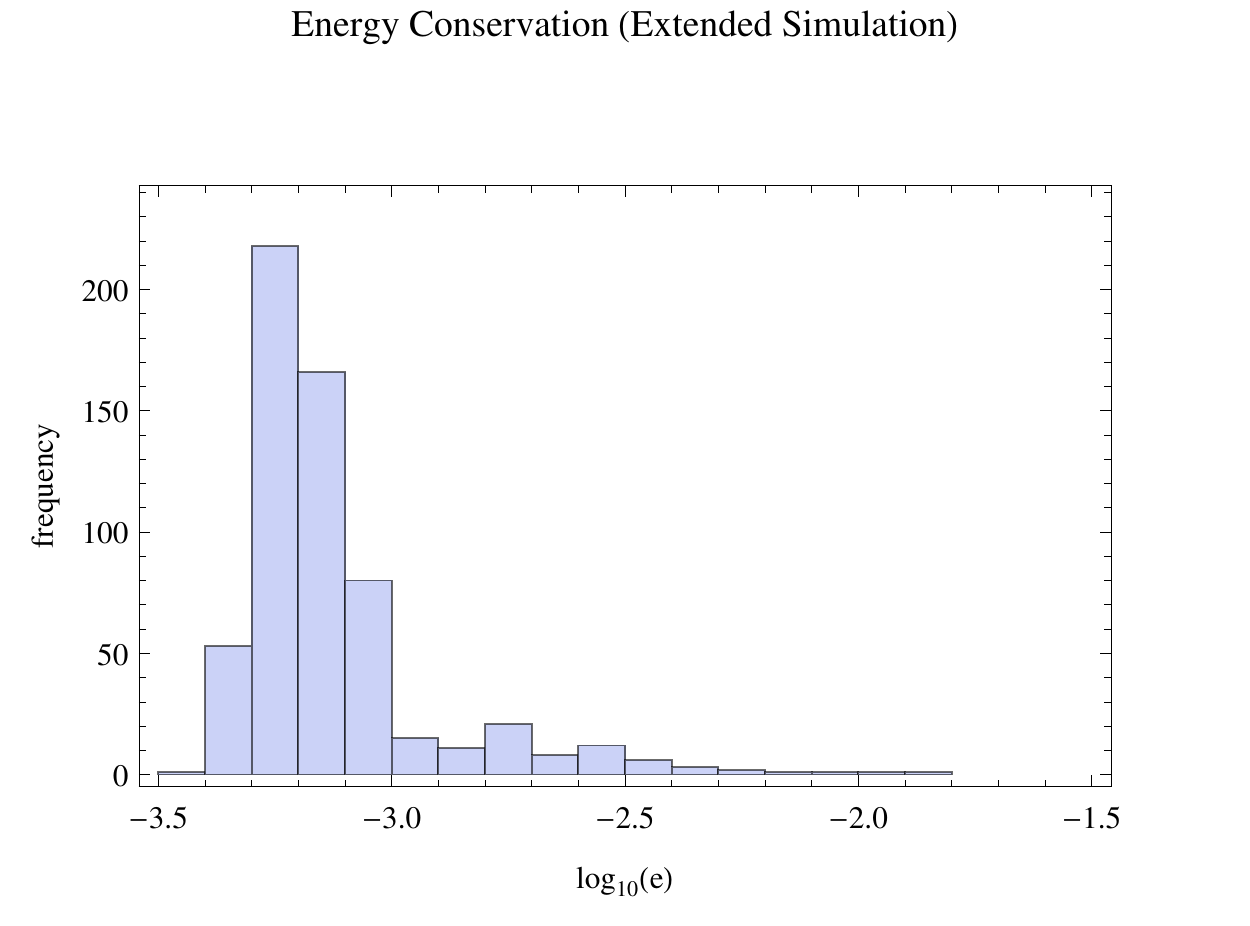}
\includegraphics[width=\columnwidth]{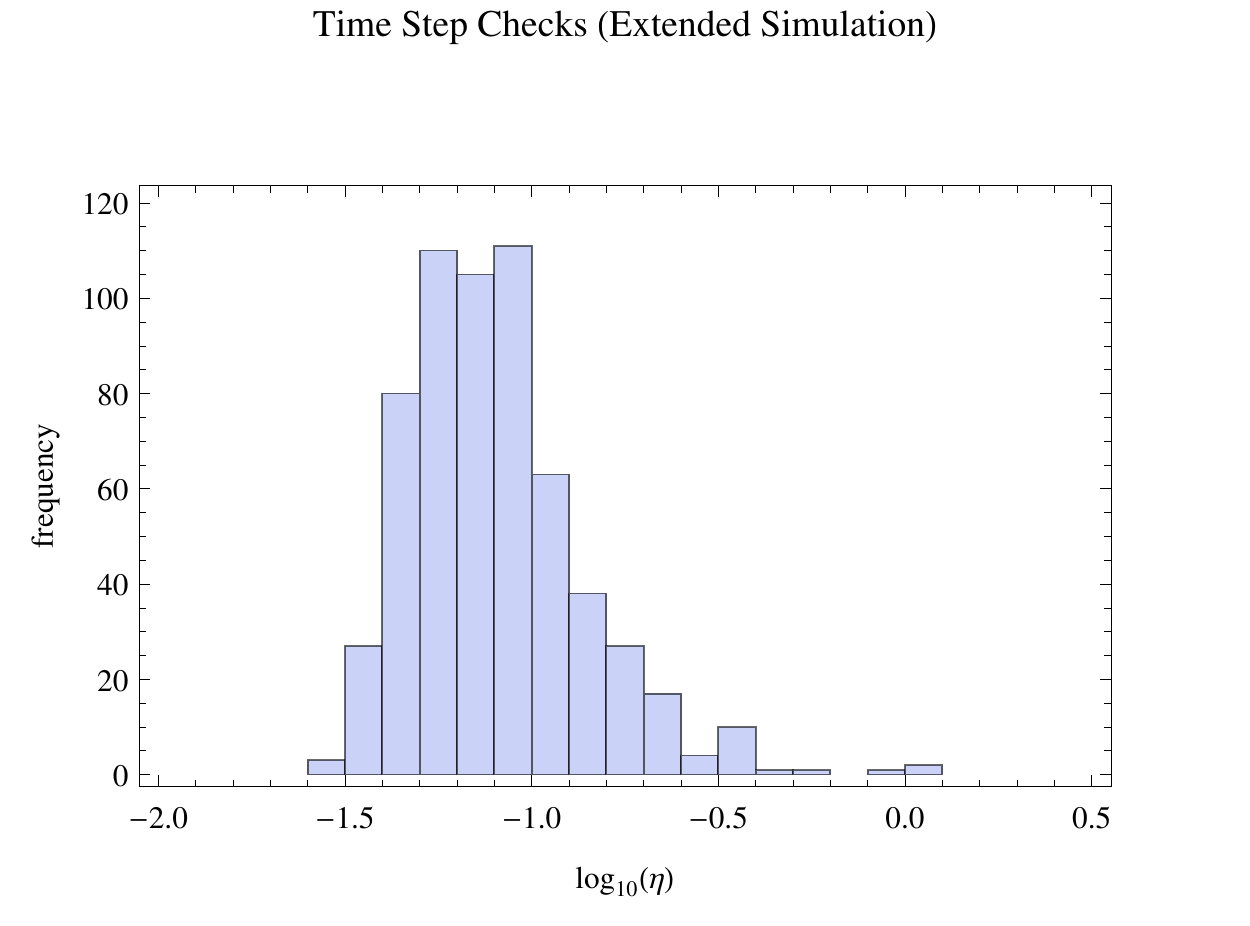}
\includegraphics[width=\columnwidth]{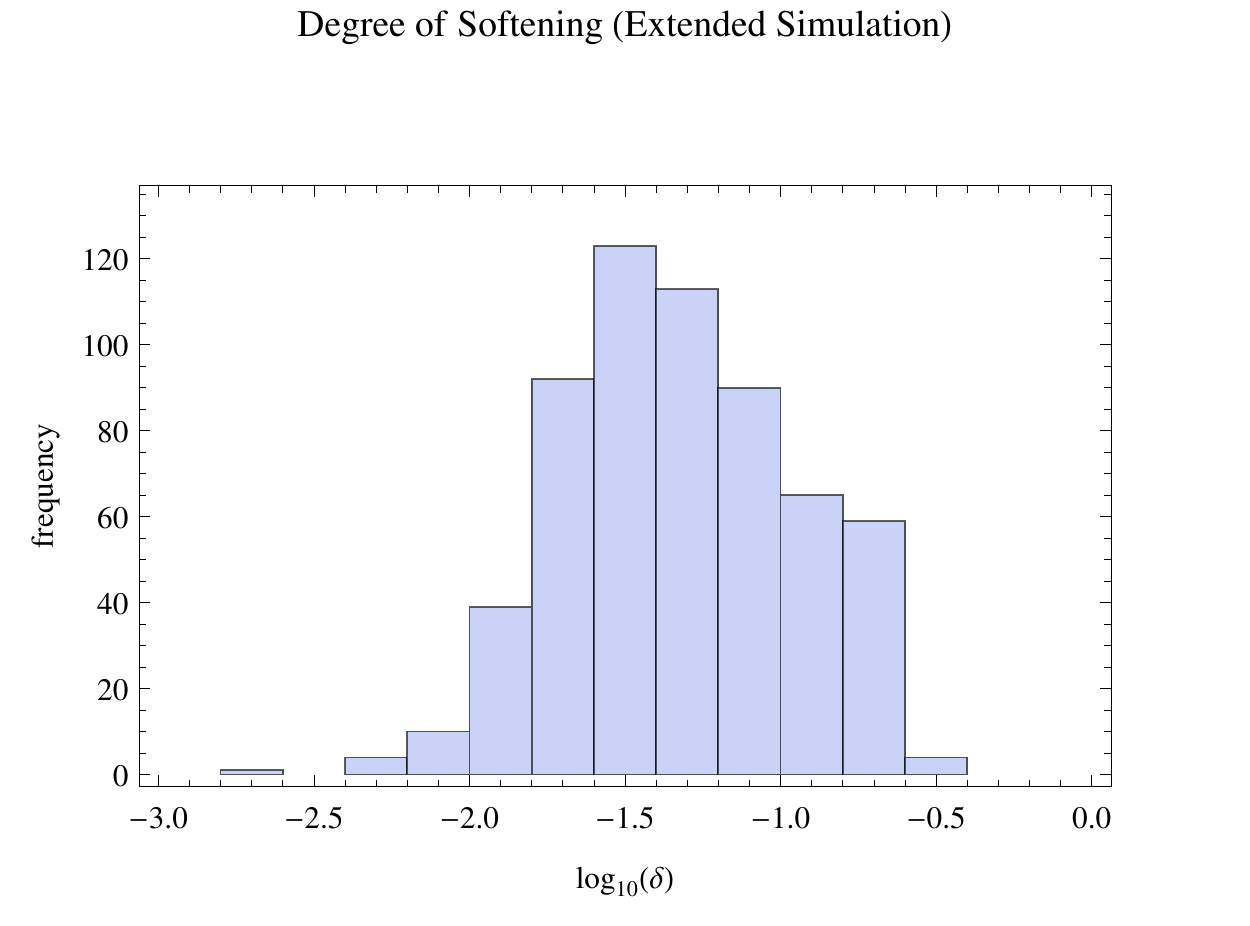}
\caption{These histograms show the distributions of logarithms of the variables we use to check internal consistency of our simulations.}
\label{fig:checks}
\end{figure}

In order to determine if our simulation parameters, such as $\Delta t$ and $\epsilon$, were suitably chosen, we keep track of three observables:
\begin{align}
\delta &\equiv \min(\Delta r/\epsilon)\\
\eta &\equiv \max\left(\frac{\Delta \vec{v}  . \Delta \vec{r}}{\Delta \vec{r}.\Delta \vec{r}} \Delta t\right)\\
e &\equiv \frac{E_f}{E_i}-1
\end{align}
The first observable $\delta$ shows the degree to which smoothing plays a role in the simulation. When $\delta$ is small there are particles in the simulation that move very close to each other. Although this is not a priori wrong, we would like to know if this behavior is systematic or not. The second variable $\eta$ represents the maximum relative distance by which two particles move towards each other per time step during the course of the simulation. If for any two particles $\eta > 1$ the simulation makes too large time steps and may miss some of the detailed dynamics. Finally the third variable $e$ represents the fractional change in the total energy of the system. Since our results are based on energies of particles in the simulation we require that $e$ be as small as possible. Figure~\ref{fig:checks} shows the distribution of values of these variables for across all 600 individual simulations in the Benchmark set. 

You can see that in the rare cases the total energy fluctuates by up to $1\%$, however, this is the case for only a few simulations and this is the most extreme deviation during each simulation. Similarly, only a few simulations contain time step during which a particle moved far compared to its distance to any other particles, as can be seen from the distribution of the $\eta$ variable. Finally, we see that particles do get within the softening distance $\epsilon$ because the minimum value of $\delta$ reached down to $10^{-2}$ in some simulations.

\subsection{Dependence on number of particles in the simulation}
\begin{figure}
\includegraphics[width=\columnwidth]{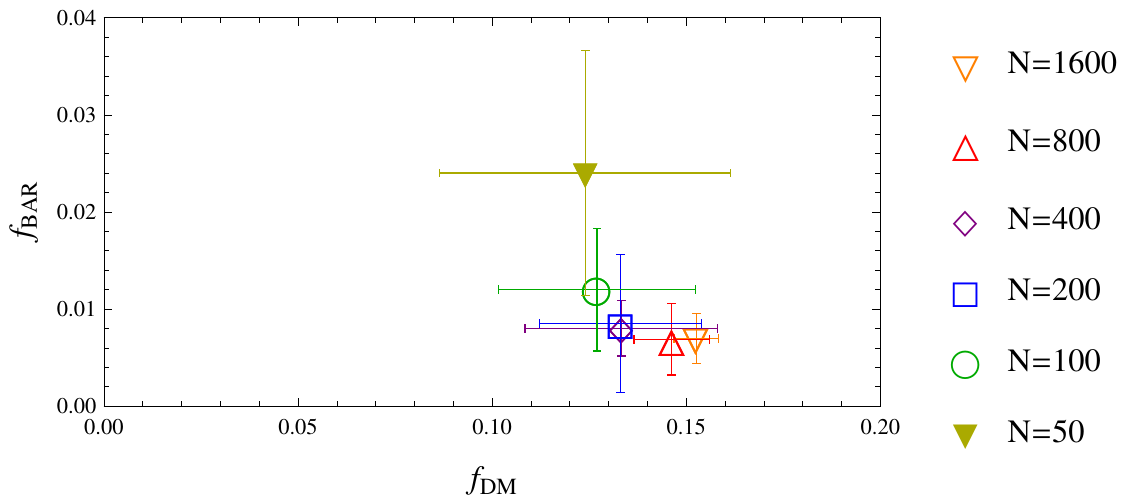}
\includegraphics[width=\columnwidth]{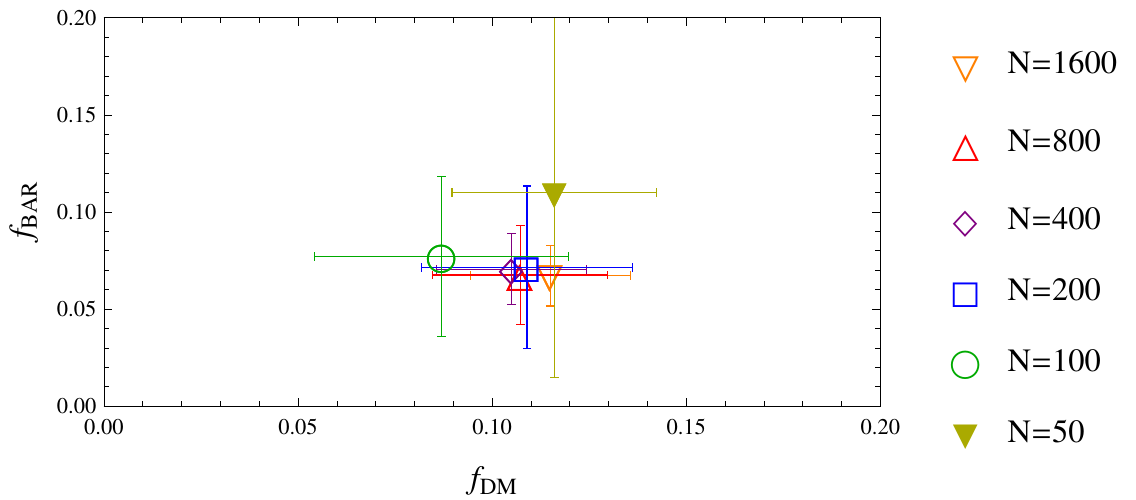}
\caption{Convergence test: Although the fraction of bound particles of each kind shows a mild dependence on the total number of particles used in the simulations, these variations are well within the error bands we derive for these quantities. The results are shown for two different scenarios.}
\label{fig:conv}
\end{figure}

Naturally, we would like our results to be independent of the number of particles we use to simulate the continuous distributions of baryons and matter. We verify this by running a series of simulations with identical parameters but different total numbers of particles $N \in \left\{50,100,200,400,800,1600\right\}$. We process each simulation and plot the resulting fractions of bound dark matter and baryons in figure~\ref{fig:conv}. Although the fraction of bound particles of each kind shows a mild dependence on the total number of particles used in the simulations, these variations are well within the error bands of each other as soon as $N \geq 100$.

\section{Simulations}

\subsection{Set 1: The Benchmark Set}

We are interested in determining how the mass of the final state and the dark matter baryon ratio depend on $x$, the size of the initial patch, and $h_D$, the thickness of the dark matter disk. Therefore, in this set of simulations we have fixed all other parameters:
\begin{align*}
R &=4\kpc\\
\epsilon &=0.02\kpc\\
N &=1024\\
\Sigma_{dm} &=10^7 M_\odot/\kpc^2\\
\Sigma_{bar} &=5\times10^7 M_\odot/\kpc^2\\
h_B &=0.3\kpc\\
v_0 &=180\kpc/\text{Gyr}
\end{align*} 
and scanned through the sets
\begin{align*}
h_{D}/\kpc &\in \{0.01,0.02,0.03,0.04,0.05,0.06\}\\
x/\kpc &\in \{0.3,0.4,0.5,0.6,0.8,1.0\}
\end{align*}
We have chosen the lower limit on the patch size $x=0.3\kpc$ because this is the scale of the Toomre instability for $h_D=0.01\kpc$. Moreover, the scale-height of baryons is also $h_B=0.3\kpc$ and so it seems unphysical to pull out any smaller patches. We stopped the simulation after $\Delta t = 2.5 \text{Gyr}$ and determined the number of bound DM particles and bound baryon particles by the method described in section~\ref{sec:simulation}. The results show that indeed the bound states contain a high fraction of dark matter -- in general dominating over the baryonic component. We find that both final dark matter to baryon ratio and final mass of the resulting clumps depend on the size of the initial patch for reasons we now explain. We have ran the same analysis after $\Delta t = 0.5 \text{Gyr}$ and the results were very similar. This makes us believe we are probing an asymptotic final state and that the final clumps we observe have virialized. 

\begin{figure}
\includegraphics[width=\columnwidth]{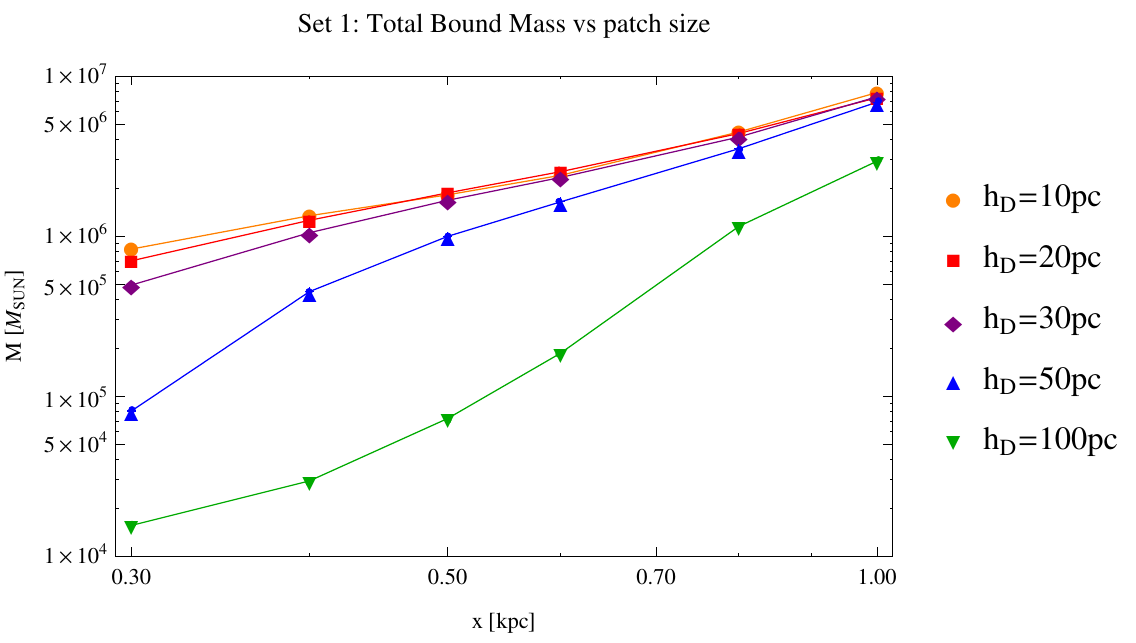}
\caption{Dependence of the final mass on the size of the initial patch.}
\label{fig:masslongx}
\end{figure}

First, we study the dependence of the final clump mass on the size of the initial patch. Naively, we expect that since the DM component is very cold, most of it stays bound, while the baryons are very hot and interacting and therefore many of them evaporate. Therefore, we predict that the mass of the final clump is close to the initial mass of the DM component and so follows the relation $m = \Sigma_{dm} x^2$. Indeed, figure~\ref{fig:masslongx} confirms this. This relation breaks down for thicker disks. In those cases the Jeans mass of the clumps becomes larger than their actual mass and the clumps are not stable, losing a significant portion of their dark matter component to evaporation.

\begin{figure}
\includegraphics[width=\columnwidth]{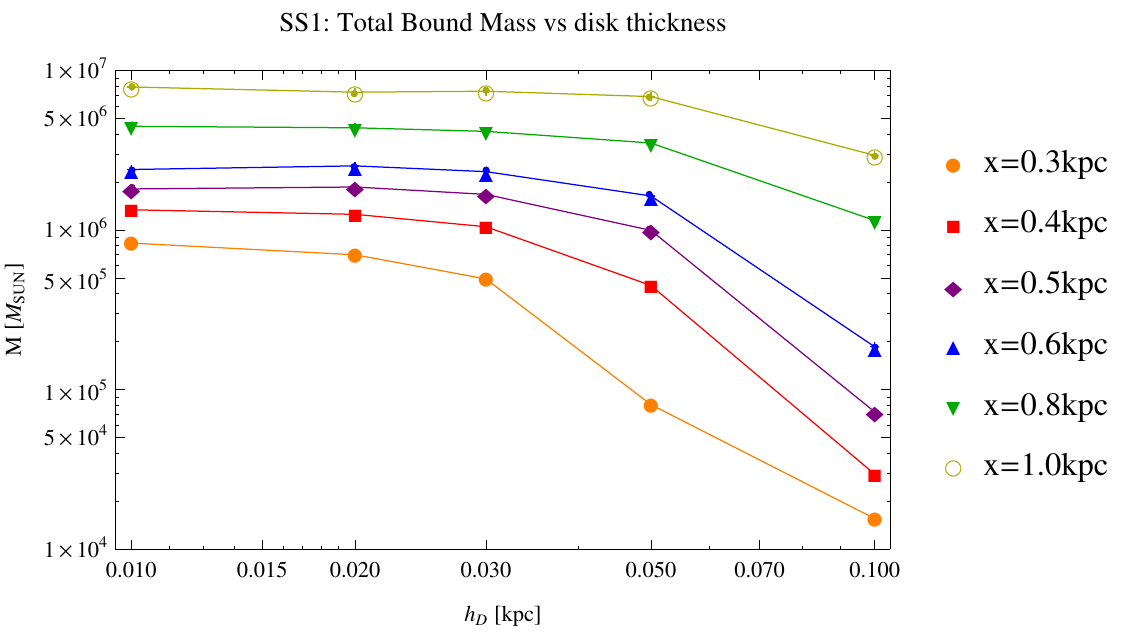}
\caption{Dependence of the final mass on the thickness of the dark disk.}
\label{fig:masslongh}
\end{figure} 
\begin{figure}
\includegraphics[width=\columnwidth]{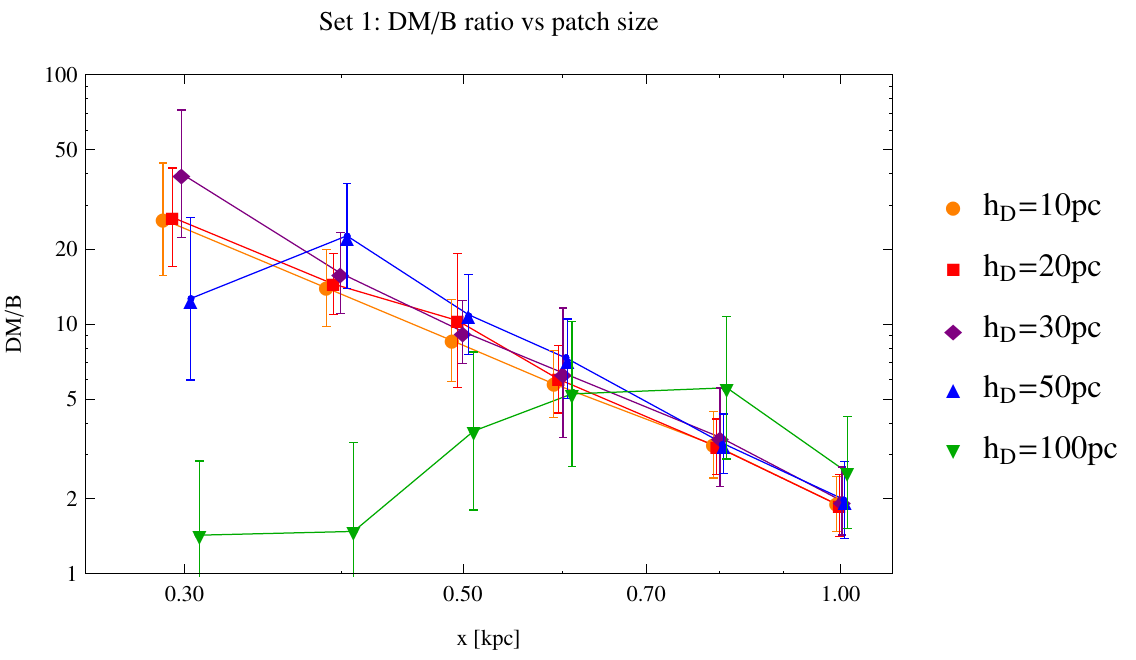}
\caption{Dependence of the final DM to baryon ratio on the size of the initial patch.}
\label{fig:ratiolongx}
\end{figure} 
\begin{figure}
\includegraphics[width=\columnwidth]{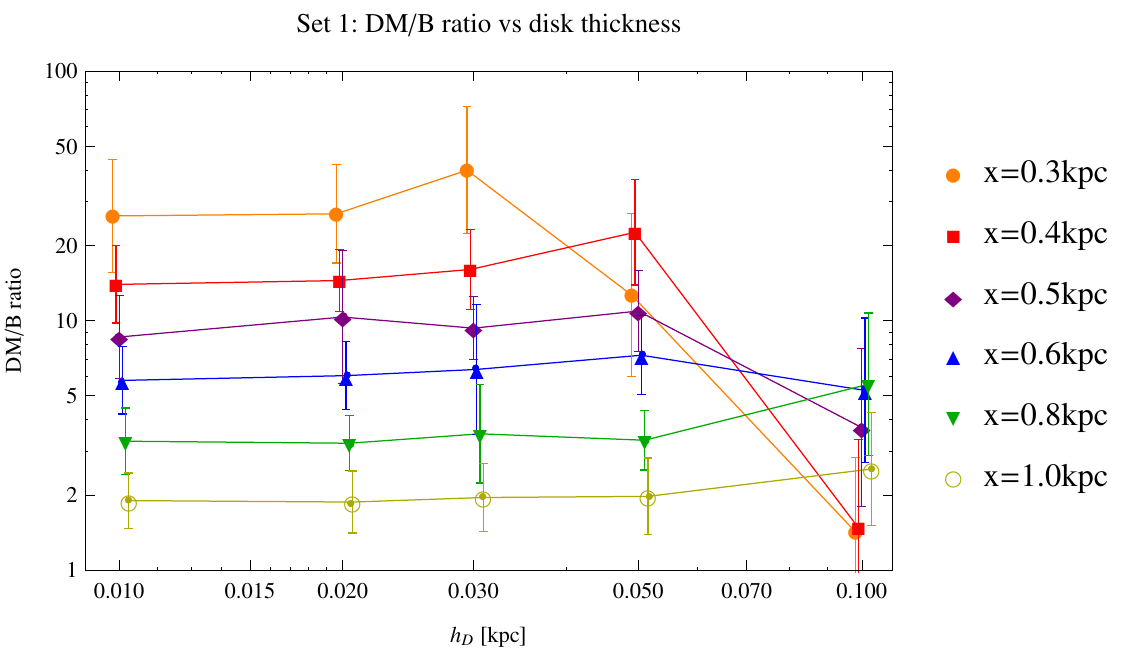}
\caption{Dependence of the final DM to baryon ratio on the thickness of the dark disk.}
\label{fig:ratiolongh}
\end{figure} 
The final ratio of dark matter to baryons shows strong dependence on the size of the initial patch. As the patch size $x$ increases the amount of angular momentum in this patch increases as $x^4$. As a result the dark matter distribution inside a large patch has a lot of angular momentum and forms a bar. This bar throws away the components with the largest specific angular momentum leaving behind a low angular momentum core. During the formation of the bar, before its re-collapse, the dark matter particles are much closer in velocities to the baryons. Since there are many more baryons in the phase space occupied by the dark matter we expect lower dark matter to baryon ratio of the final bound state. This can be seen in figures~\ref{fig:ratiolongx}~and~\ref{fig:ratiolongh}. 

Our simulations show that PIDM can produce clumps of matter and hence possibly dwarf satellite galaxies with a highly increased dark matter to baryon ratio from tidal interactions between merging galaxies. Because this ratio is highly dependent on the scale of the objects that are pulled out by the tidal forces,  a dedicated full simulation of this event is essential to determining the dark matter to baryon ratio of the final bound object.

\subsection{Set 2: Increased Mass Set}

We have repeated our analysis with the same parameters but with increased mass of the Andromeda Galaxy by a factor of $1.5$. The SPG is then the same as the Milky Way scaled down by a factor of $0.375$. This  changes the characteristic length scales and rotational velocities as we have indicated in section~\ref{sec:progenitor}. The patches we investigated were still pulled out from $R=4.0\kpc$ from the center of the smaller progenitor galaxy in order to be able to compare to scenarios from the Benchmark Set. We find that the change in the dark matter baryon ratios were not statistically significant. However, we observe that the total bound mass increased by a factor of $2-3$ compared to the the Benchmark Set. When we rescaled Andromeda mass by a factor of $1.5$, the radial scale factor of the disk increased by a factor of $\sqrt{1.5}$. Therefore, at $R=4.0\kpc$ the density increased by a factor of $\sim 1.65$. This means that the increase in mass was not linear, and the essential non-linearity of this gravitational system became apparent.

\subsection{Set 3: Different Location in the Disk}

We performed another study in which we changed the location of the original patch while keeping all other parameters fixed. We chose a patch one scale radius further away from the center of the SPG -- at $R=5.5\kpc$, as opposed to the original $R=4.0\kpc$ used in the Benchmark Set. We did not explore any patches closer to the center of the SPG, because these patches are much less likely to get pulled out (objects closer to the center are more strongly bound). Similarly, we did not chose patches any further from the center of the SPG because the final masses of the resulting bound objects become too small. However, note that if more than one patch get pulled out of the SPG and subsequently merge they can still form massive objects.

We find that the final dark matter to baryon ratios are not statistically different between the $R=4.0\kpc$ and $5.5\kpc$ scenarios. While we expect the final bound masses to be at least $e^{-1}$ lower, they in fact turn out to be lower by a factor of $\sim 5$, most likely due to non-linearity of the system just as it was the case in set 2.

\subsection{Set 4: Fixed Mass Set}
Our previous simulations have shown a strong dependence on the size of the initial patch. However, in simulations with fixed number of particles and surface density, increasing the patch size means that each particle has to have a larger mass. In order to check that this scaling is not an artifact of our choices of particle mass we set up another set of simulations. In this set we changed the number of particles in proportion to the total simulated mass with every other variable fixed as in the previous case. Table~\ref{tab:tfixed} shows the values of $N$ we used in order to fix the mass of each particle to $10^4 M_\odot$.
\begin{table}
\begin{center}
\begin{tabular}{ccc}
\hline 
$x [kpc]$ & $N(\text{Extended Run})$ & $N(\text{Fixed Mass Run})$ \\ 
\hline 
0.3 &1024& 540 \\ 
0.4 &1024& 960 \\ 
0.5 &1024& 1500 \\ 
0.6 &1024& 2160 \\ 
0.8 &1024& 3840\\
\hline
\end{tabular}  
\end{center}
\caption{Values of $N$ for different patch sizes for both the Extended Run and the Fixed Mass Run.}
\label{tab:tfixed}
\end{table}

\begin{figure}
\includegraphics[width=\columnwidth]{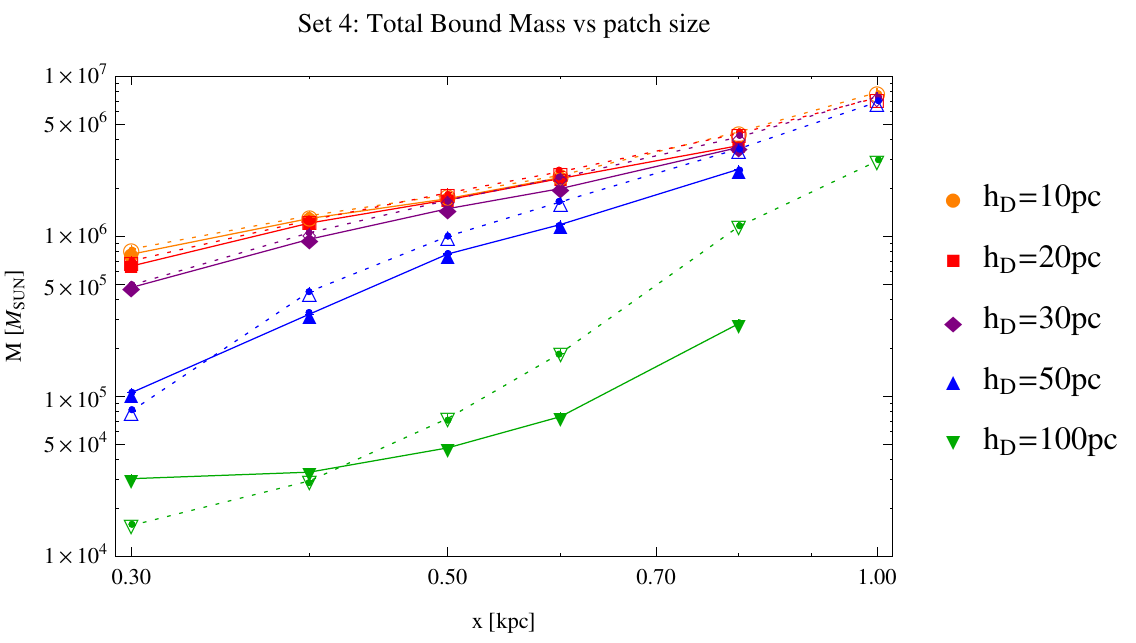}
\caption{Dependence of the final mass on the size of the initial patch. The results from the Benchmark Set are dotted. }
\label{fig:massfixedx}
\end{figure}

\begin{figure}
\includegraphics[width=\columnwidth]{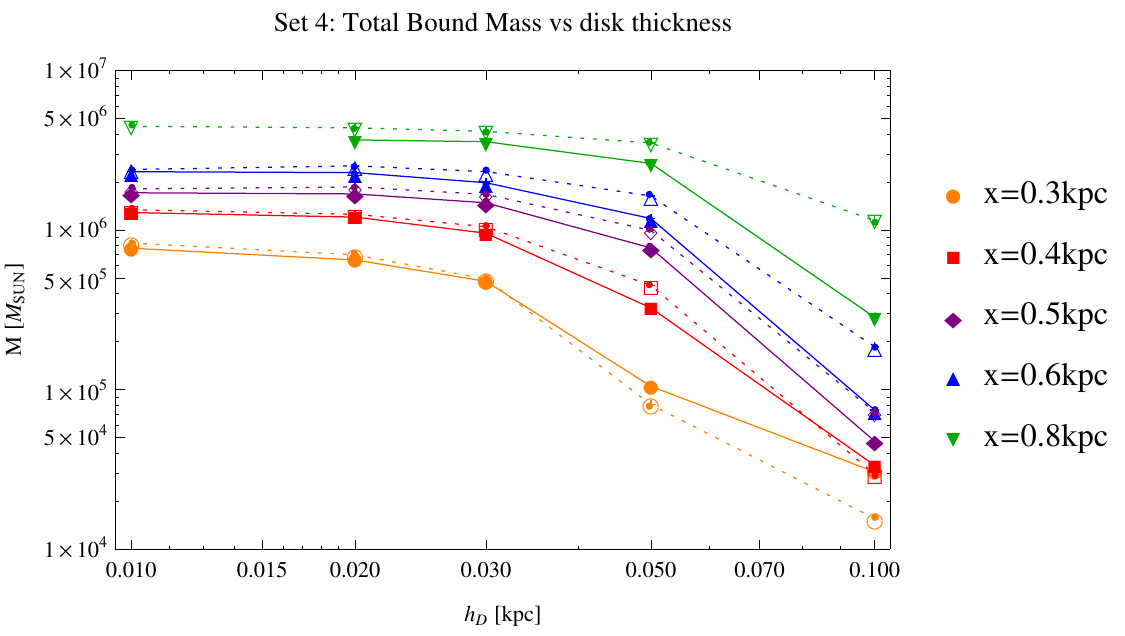}
\caption{Dependence of the final dark matter mass on the thickness of the dark disk. The results from the Benchmark Set are dotted.}
\label{fig:massfixedh}
\end{figure}

\begin{figure}
\includegraphics[width=\columnwidth]{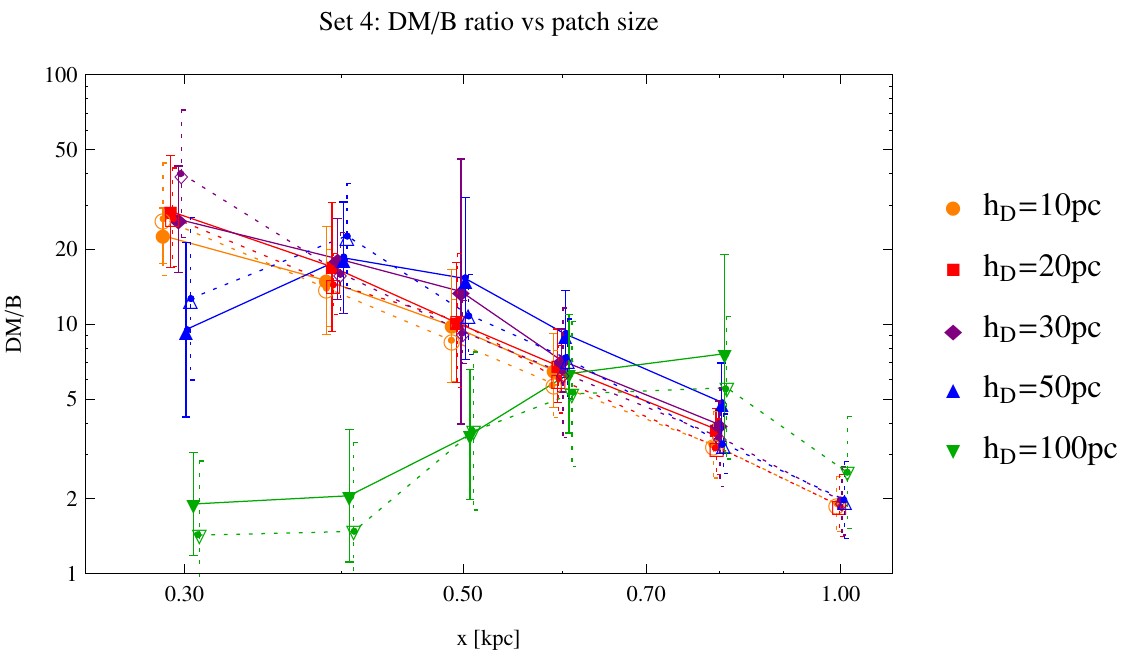}
\caption{Dependence of the final DM to baryon ratio on the size of the initial patch. The results from the Benchmark Set are dotted. We have added a small offset to the $x$ values for different series in order to allow the reader to resolve all the error bars. }
\label{fig:ratiofixedx}
\end{figure}

\begin{figure}
\includegraphics[width=\columnwidth]{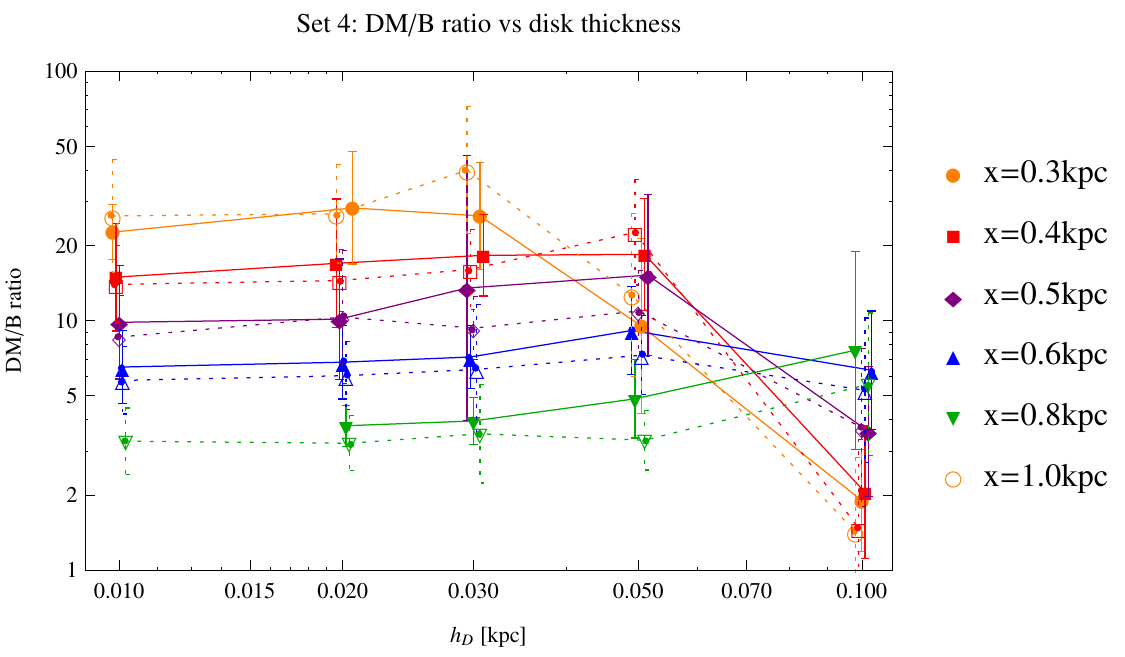}
\caption{Dependence of the final DM to baryon ratio on the thickness of the dark disk. The results from the Benchmark Set are dotted. We have added a small offset to the $h_D$ values for different series in order to allow the reader to resolve all the error bars.}
\label{fig:ratiofixedh}
\end{figure}

The results are very similar to the Benchmark Set and we do not see any systematic deviations, as can be seen from figures~\ref{fig:massfixedx},~\ref{fig:massfixedh},~\ref{fig:ratiofixedx} and \ref{fig:ratiofixedh}. 

\section{Conclusion}

In this paper we have investigated a possible explanation for the planar distribution of satellite dwarf galaxies with high dark matter content. We find that presence of a thin dark disk formed from cooled dissipative dark matter during galactic collisions can result in formation of DDDM rich tidal dwarf galaxies after galactic mergers. In our simulations, the final dark matter to baryon ratio could reach as high as $\sim 30$. Such numbers are in agreement with some of the observed $M/L$ values for the tidal dwarf in the plane around Andromeda galaxy. However, this ratio depends on the size of the region that gets ejected during the merger of the two galaxies. Our study does not determine the scale of these patches and further full simulations of galactic mergers containing dark disks have to be performed in order to answer this question.

The experimental values of $M/L$ show a significant range and our model could account for this by identifying different final values of $M/L$ with different sized patches or patches that were pulled out from different positions in the disk of one of the proto-galaxies. A full N-body simulation of the merger would allow us to predict the distribution of $M/L$ values for these tidal galaxies and put additional constraints on our model. It will be interesting when measurements and N-body simulations advance to the point we will be able to constrain the parameters of these models.

We expect that out-of-the-plane satellites have large $M/L$ values. At this point, it seems coincidental that they are not statistically different from in-the-plane satellites in terms of their $M/L$.

There are several potentially testable consequences of this model as well as exciting questions in need of answer: 

We predict that these tidal dwarf galaxies are dominated by their DDDM component. As a result their density profiles may deviate from the CDM and baryonic density profiles we would expect to find in these tidal dwarf galaxies. We are currently investigating what shape these profiles take in order to allow future comparison with experimental data and distinguish PIDM dominated galaxies from ones with high usual CDM content. 

Another observational strategy could be to search for radiation from these objects. We do not expect any annihilation signals from these DDDM rich tidal dwarf galaxies, because the DDDM was treated as asymmetric. However, if there is non-zero kinetic mixing between the ordinary and dark photons, it may be possible to detect radiation from the excitations of the bound states in DDDM and a detailed study is necessary to investigate the potential constraints on these models.

It will also be of interest to study further the metallicity of the dwarf galaxies. In present day galaxies we observe a correlation between metallicity of gas clouds and their temperature (higher metallicity allows faster cooling to lower temperatures). Since we also observe a correlation between velocity distributions (temperature) and baryon abundance in the tidal dwarf our scenario predicts tidal dwarfs with higher metallicity should contain larger baryonic component as can be observed in the dwarf galaxies. However, a more quantitative statement can be only obtained after detailed full simulation which tracks metallicity of the baryonic gas. 
  
In conclusion, our work illustrates the possibility of achieving high mass to luminosity ratios in tidal dwarf galaxies. This mechanism is unique to the PIDM scenario, which has just the right properties to allow for low velocity dark matter that can bound into satellites. A  dedicated full N-body simulation of galactic mergers with embedded DDDM disks can help confirm these predictions. 

\section{Acknowledgements} We are grateful to Matthew Walker for alerting us to the issue of the planarity of dwarf galaxies in Andromeda and other galaxies and for many helpful discussions on the subject. We would also like to thank Scott Tremaine, Lars Hernquist, James Guillochon, Shy Genel and especially Matt Reece for useful discussions and comments on our manuscript. The work of LR was supported in part by NSF grants PHY-0855591 and PHY-1216270. JS would like to thank the Center For Fundamental Laws of Nature Initiative.

\appendix

\section{Plummer Sphere}
\label{sec:plummer}
In order to model an extended distribution of mass we use the Plummer Sphere\cite{Plummer}. The mass density distribution for this model is:
\begin{equation}
\rho(r,\epsilon) = \frac{3M}{4\pi}\frac{\epsilon^2}{\left(r^2 + \epsilon^2\right)^{5/2}} 
\end{equation}
This gives simple expressions for the potential and force between the two particles:
\begin{align}
\Phi(r) &= -\frac{GM^2}{\left(r^2+\epsilon^2\right)^{1/2}}\\
\vec{F}(r) &= -\frac{GM^2 \vec{r}}{\left(r^2+\epsilon^2\right)^{3/2}}\\
\end{align}
Notice that the potential and force reduce to their Newtonian expressions in the limit $\epsilon \rightarrow 0$. However, the choice $\epsilon \neq 0$ regulates the singularity at $r=0$.  

\section{Leapfrog Algorithm}
\label{sec:leapfrog}
In order to numerically integrate the equations of motion we have to choose a particular integration method. We chose a relatively simple Leapfrog algorithm. This algorithm works with positions at integers steps and velocities at half-integers steps, which is where its name comes from. Schematically it can be written down as:
\begin{align*}
x_i &= x_{i-1}+v_{i-1/2} dt \\
a_i &= F(x_i)/m \\
v_{i+1/2} &= v_{i-1/2} + a_i dt \\
x_{i+1} &= x_i + v_{i+1/2} dt \\
\vdots
\end{align*}
where $F(x)$ is the net force acting on the particle. In our case it is the sum of the gravitational forces from all other particles in the simulation.

\nocite{*}
\bibliographystyle{utphys}
\bibliography{VTDSbib}

\end{document}